\documentclass[preprint,floatfix,amsmath,amssymb,aps,superscriptaddress]{revtex4-2}
\usepackage{graphicx}
\usepackage{bm}
\usepackage{booktabs}
\usepackage{xcolor}
\usepackage{float}
\usepackage{hyperref}

\renewcommand{\textcolor}[2]{#2}

\begin{document}

\title{\textcolor{red}{Spin Splitter without Spin-Split Bands:
A Reconfigurable Altermagnetic Texture}}

\author{Bin Xi}
\email{binxi@yzu.edu.cn}
\affiliation{College of Physics Science and Technology, Yangzhou University, Yangzhou 225002, China}

\author{Jie Lu}
\affiliation{College of Physics Science and Technology, Yangzhou University, Yangzhou 225002, China}

\author{Qiang Luo}
\affiliation{College of Physics, Nanjing University of Aeronautics and Astronautics, Nanjing, 211106, China}

\author{Ken Chen}
\affiliation{Lanzhou Center for Theoretical Physics, Lanzhou University, Lanzhou 730000, China}
\affiliation{Key Laboratory of Quantum Theory and Applications of MoE, Lanzhou University, Lanzhou 730000, China}
\affiliation{Key Laboratory of Theoretical Physics of Gansu Province $\&$ Gansu Provincial Research Center for Basic Disciplines of Quantum Physics, Lanzhou University, Lanzhou 730000, China}
\affiliation{School of Mathematics and Physics, Southwest University of Science and Technology, Mianyang 621010, China}

\author{Jia-Wei Mei}
\affiliation{Department of Physics, Southern University of Science and Technology, Shenzhen 518055, China}

\author{Hong-Gang Luo}
\affiliation{Lanzhou Center for Theoretical Physics, Lanzhou University, Lanzhou 730000, China}
\affiliation{Key Laboratory of Quantum Theory and Applications of MoE, Lanzhou University, Lanzhou 730000, China}
\affiliation{Key Laboratory of Theoretical Physics of Gansu Province $\&$ Gansu Provincial Research Center for Basic Disciplines of Quantum Physics, Lanzhou University, Lanzhou 730000, China}

\author{Jize Zhao}
\email{zhaojz@lzu.edu.cn}
\affiliation{Lanzhou Center for Theoretical Physics, Lanzhou University, Lanzhou 730000, China}
\affiliation{Key Laboratory of Quantum Theory and Applications of MoE, Lanzhou University, Lanzhou 730000, China}
\affiliation{Key Laboratory of Theoretical Physics of Gansu Province $\&$ Gansu Provincial Research Center for Basic Disciplines of Quantum Physics, Lanzhou University, Lanzhou 730000, China}

\date{\today}

\begin{abstract}
The altermagnetic spin-splitter effect converts an electric field into a
transverse pure spin current, with no net magnetization and no charge-Hall
counterpart.  In established materials this function is tied to crystal-fixed
spin-split bands that lock the polarization axis to the
\textcolor{red}{lattice}.  We show that \textcolor{red}{the noncoplanar
counter-spiral ground state of a frustrated honeycomb magnet instead carries
the altermagnetic operation through a $\mathbf Q$-locked helicity
mirror $g$.  The mirror selects the spin-current polarization and forbids the
perpendicular one, while an antitranslation $\Theta$ forbids even-parity
spin splitting.}  Band splitting and spin-splitter response therefore rest on
different symmetry elements.  \textcolor{red}{Either element alone enforces
the charge-Hall zero---a redundancy absent from other spin--orbit-free
noncollinear routes---and a charge Hall appears only when both elements are
removed.}  Hole doping then
realizes a
\emph{spin splitter without spin-split bands}\textcolor{red}{---the
symmetry-allowed odd-parity residual below
$2\times10^{-7}$ of the hopping $t$ at the Fermi level---with
$\sigma_H^{(s_y)}=0.082\,e^2/h$} without spin--orbit coupling
and with zero charge Hall response.  Selecting among the three degenerate
$\mathbf{Q}$ orientations rotates the polarization axis in exact $120^\circ$
steps \textcolor{red}{at fixed magnitude and charge-Hall zero; the
selection rules persist in
a $32$-site cell accessible to programmable photonic and circuit lattices.}
\end{abstract}

\maketitle

\paragraph{Introduction.---}%
Altermagnets~\cite{SmejkalBeyond2022,Smejkal2022,Hayami2019SpinSplitting,
Yuan2020GiantSpinSplitting,Yuan2021LowZMaterials,NatRevAltermagnetFunctional2025,
KitaevBilayer,StaggeredDMcanting2026}
are compensated magnets that combine vanishing net magnetization with
spin-polarized electronic responses, enabled by a spin-space-group (SSG)
element $[S_\text{spin}\!\parallel\!R_\text{real}]$ other than pure
translation or inversion~\cite{BrinkmanElliott1966,LitvinOpechowski1974,
JiangSpinSpaceGroups2024,ChenSpinSpaceGroups2024,XiaoSpinSpaceGroups2024,
WatanabeSpinCrystallographic2024,LiuOSSG2026}\textcolor{red}{, a definition
that extends to noncollinear order~\cite{CheongHuang2024noncollinear}}.
Their principal device application is the \emph{spin-splitter}
effect~\cite{GonzalezHernandez2021SpinSplitter,SmejkalHall2020,
Ma2021PiezomagnetismSpinCurrent,NatRevAltermagnetFunctional2025}: an applied electric field
generates a transverse pure spin current without a charge-Hall counterpart and
without ferromagnetic stray fields.  In established collinear materials such
as MnTe and CrSb, this function is tied to crystal-fixed spin-split
bands~\cite{SmejkalHall2020,GonzalezHernandez2021SpinSplitter,
Osumi2024MnTe,Krempasky2024MnTe,Lee2024MnTe,Reimers2024CrSb,Ding2024CrSb,
MnTeOrbital2025}, so the spin-current polarization direction
is locked to the crystallographic frame.
\textcolor{red}{Because one crystal-fixed operation produces both the
splitting and the response, resolved band splitting has become the practical
screening signature for collinear altermagnets.  The spectral and transport
roles, however, need not rest on the same symmetry element.  Noncollinear
routes such as
Mn$_3$X and antiferromagnetic skyrmion crystals already give
spin--orbit-free transverse spin currents with no
charge Hall.  In those cases the charge-Hall zero rests on one structural
condition: coplanarity or equivalent
sublattices~\cite{Zhang2018NoncollinearSHE,Gobel2017AFMSkX}.  We show that the
counter-spiral instead carries an independent non-translation SSG element,
which supplies the spin-splitter selection rules without even-parity
splitting and protects the charge-Hall zero together with the
antitranslation.}

\begin{figure}[H]
  \includegraphics[width=0.62\textwidth]{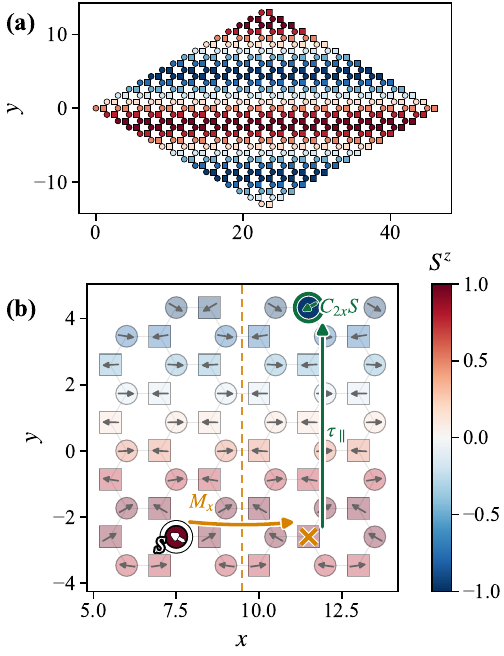}
  \caption{Representative counter-spiral texture.  The two sublattices carry
  single-$\mathbf Q$ helices of opposite handedness.  (a) $S^z$ map of the
  $16\!\times\!16$ magnetic supercell (512 sites); circles/squares mark the
  $A/B$ sublattices.  (b) Local zoom: site color is $S^z$ and arrows are
  $(S^x,S^y)$.  \textcolor{red}{The path shows} the spatial Seitz part of the helicity mirror
  $[C_{2x}\!\parallel\!\{M_x|\bm{\tau}_g\}]$ in reduced form---reflection about
  the glide line $x\!=\!19/2$ (dashed orange), then
  $\bm{\tau}_\parallel\!=\!(0,4\sqrt{3})$.  \textcolor{red}{White $A$ (spin $S$)
  maps through the yellow waypoint to green $B$ (spin $C_{2x}S$).}}
  \label{fig:texture}
\end{figure}

\textcolor{red}{The noncoplanar counter-spiral of Fig.~\ref{fig:texture} is
the platform.}  It
is generated by the classical honeycomb $J$-$\Gamma'$-$D$ model of the
Supplemental Material~\cite{SMnote} [Sec.~S1] and motivated
by bond-anisotropic exchanges in high-spin van der Waals honeycomb
magnets~\cite{Xu2018CrKitaevSIA,Xu2020Spin32Kitaev,Stavropoulos2021CrI3Anisotropy}.
Related counterrotating orders are also reported in Li$_2$IrO$_3$
polytypes~\cite{Williams2016alphaLi2IrO3,Biffin2014betaLi2IrO3,Biffin2014gammaLi2IrO3,KimchiColdeaVishwanath2015}.
\textcolor{red}{The texture itself carries two exact spin-space-group
elements: an \emph{antitranslation} $\Theta$ (half-period spin reversal) and a
$\mathbf{Q}$-locked \emph{helicity mirror} $g$.  Bands and Hall responses
appear only once the texture is coupled to itinerant electrons, which we do
through a spin--orbit-free $s$--$d$ Hamiltonian.  In that coupled problem the
two elements act on different observables.  $\Theta$ forbids
even-parity band splitting and the charge Hall response.  $g$ is the
non-translation altermagnetic operation: it fixes the allowed Hall spin
channel and its
polarization axis, forbids the perpendicular spin channel (the \emph{dark
axis}), and independently forces
the charge Hall to vanish at every filling.  Thus $\Theta$ alone protects the
even-splitting zero, $g$ alone protects the dark axis, and either element
alone protects the charge-Hall zero: removing one leaves it intact, and only
removing both makes a charge Hall symmetry-allowed.}

\textcolor{red}{In the intact texture, hole doping realizes a \emph{spin
splitter without spin-split bands}---the even-parity channel forbidden by
$\Theta$, the symmetry-allowed odd-parity residual below $2\times10^{-7}\,t$
at the Fermi level---with
$\sigma_H^{(s_y)}=0.082\,e^2/h$ and zero charge Hall, hence a pure transverse
spin current under electric drive.  The separate breaking tests leave the
$s_y$ spin-Hall component essentially unchanged:
$\sigma_H^{(s_y)}$ varies by at most $0.51\%$ when $\Theta$, $g$, or both are
removed [Sec.~S6.1].  Selecting among the three
degenerate $\mathbf Q$ orientations rotates the polarization axis in exact
$120^\circ$ steps.  The $\Theta$-broken configuration isolates $g$ as the
sole remaining symmetry enforcing the charge-Hall zero: the
mirror-compatible spin--orbit term tested in
Sec.~S6.2 preserves $g$, and the zero persists.}

\paragraph{Model and texture.---}%
We consider itinerant electrons on the honeycomb lattice coupled to a preformed
classical spin texture via the $s$--$d$
Hamiltonian~\cite{MartinBatista2008,AkagiUdagawaMotome2012}
\begin{equation}
  \mathcal{H} = -t\!\!\sum_{\langle ij\rangle\sigma}\!c^\dagger_{i\sigma}c_{j\sigma}
  + J_H\sum_i c^\dagger_i(\mathbf{S}_i\!\cdot\!\bm{\sigma})c_i\textcolor{red}{.}
  \label{eq:H}
\end{equation}
We set $t=1$ as the energy unit and take $J_H=5$ throughout unless stated
otherwise.  The fixed background is the
counter-spiral texture in Fig.~\ref{fig:texture}, generated by the
classical $J$-$\Gamma'$-$D$ local-moment Hamiltonian $\mathcal{H}_{\rm loc}$
[Sec.~S1] at $J\!=\!0$, $\Gamma'\!=\!-D<0$.
This compensated $L\!=\!16$ texture has a $16\!\times\!16$ magnetic period
and contains 512 honeycomb sites, giving 1024 spinful electronic bands.
We diagonalize
$\mathcal{H}(\mathbf{k})$ on a $16\!\times\!16$ supercell BZ grid and
extract band-resolved spin polarizations
$P^\mu_n(\mathbf{k})=\langle n,\mathbf{k}|\sum_i\sigma^\mu_i|n,\mathbf{k}\rangle$.

\paragraph{Two exact spin-space-group elements.---}%
\textcolor{red}{We write spin-space-group operations as
$[S_\text{spin}\!\parallel\!R_\text{real}]$, with the full notation defined in
the Supplemental Material~\cite{SMnote}.}
The antitranslation $\Theta=[T\!\parallel\!\{E|\bm{\tau}_{1/2}\}]$ holds
because translating by half a magnetic period
($\bm{\tau}_{1/2}=8\mathbf{a}_2$ on the $L\!=\!16$ cell) reverses every moment,
$\mathbf{S}(\mathbf{r}+\bm{\tau}_{1/2})=-\mathbf{S}(\mathbf{r})$
[Sec.~S2.1].  $\Theta$ is the
symmetry class of a conventional antiferromagnet and of $p$-wave
magnets~\cite{Hellenes2023pwave,Yamada2025pwaveHelix}: it is antiunitary,
maps $\mathbf{k}\!\to\!-\mathbf{k}$, and sends a state with spin polarization
$P^\mu$ to an equal-energy partner with spin polarization $-P^\mu$ at
$-\mathbf{k}$,
so any even-parity ($d$-wave-type) spin splitting is forbidden
\emph{exactly}. Only odd-parity splitting is symmetry-allowed.
This parity constraint acts on the spectrum.  The transport
consequences below follow from the same $\Theta$ applied to the response
operators, and are independent of the splitting channel.
For the charge Hall response, which is $T$-odd, the
equal-energy partners at $\mathbf{k}$ and $-\mathbf{k}$ contribute with
opposite signs, so their charge-Hall contributions cancel at any chemical
potential, including after doping. For the spin-Hall response, the spin-current
operator combines spin and velocity. Both reverse under $T$, making this operator
$T$-even. The paired spin-Hall contributions need not
cancel, allowing a finite response.  This charge--spin distinction is the
symmetry basis for the transport phenomena discussed below.

The second element is the helicity mirror
$g=[C_{2x}\!\parallel\!\{M_x|\bm{\tau}_g\}]$, where
$M_x:(x,y)\!\to\!(-x,y)$, with glide
$\bm{\tau}_g=(19,4\sqrt{3})$ \textcolor{red}{in units of the
nearest-neighbor bond length}
[Sec.~S2.1], and $C_{2x}$ sends
$(\sigma^x,\sigma^y,\sigma^z)\!\to\!(\sigma^x,-\sigma^y,-\sigma^z)$
[Fig.~\ref{fig:texture}(b)].
This operation exchanges the two honeycomb sublattices and relates their
opposite-helicity spin textures. At the Bloch level
\begin{equation}
  \mathcal{U}_{g}(\mathbf{k})\,\mathcal{H}(\mathbf{k})\,
  \mathcal{U}_{g}^{\dagger}(\mathbf{k})
  = \mathcal{H}(M_x\mathbf{k})\, ,
  \label{eq:ssg_covariance}
\end{equation}
where $\mathcal{U}_{g}$ contains the Seitz phase and the $C_{2x}$ spin
rotation.  Geometrically, neither factor alone is a symmetry: the real-space
mirror-glide flips the spiral propagation direction and exchanges sublattices,
while the spin-space $C_{2x}$ reverses the helicity.  Because the $A$ and $B$
sublattices carry helices of opposite handedness, these changes compensate.
Equivalently,
$\mathbf{S}(M_x\mathbf{r}+\bm{\tau}_g)=C_{2x}\mathbf{S}(\mathbf{r})$,
so the combined SSG operation leaves the
counter-spiral invariant.  A single-helicity spiral on the same lattice would
fail this condition.
Unlike $\Theta$, this mirror-glide is a
\emph{non-translation} SSG operation.  \textcolor{red}{Within the generalized
noncollinear extension of altermagnetism~\cite{CheongHuang2024noncollinear},
this spin--lattice operation identifies the compensated counter-spiral as an
altermagnet.  It coexists with the antitranslation of the $p$-wave-magnet
class as a distinct symmetry element.}
In the spin--orbit-free charge sector the spin part of $g$ drops out,
leaving the orbital point group $G_{\rm orb}=\{E,M_x\}=m$: a single mirror
and no inversion [detailed in Sec.~S2.2].  This mirror makes
the charge Berry curvature odd under $M_x$ \textcolor{red}{and therefore
enforces a zero charge Hall response, while $\Theta$ independently enforces
the same zero through partner cancellation.  Their remaining constraints are
distinct: only $\Theta$ forbids even splitting, whereas only $g$ forbids the
dark spin channel.  A $g$-preserving perturbation can therefore release the
spectral constraint while retaining the charge-Hall and dark-spin zeros.
Conversely, a $\Theta$-preserving sublattice imbalance can break $g$ while
retaining the even-splitting and charge-Hall zeros, but the dark spin channel
is no longer symmetry-forbidden.  Only when both elements are removed is a
charge-Hall response symmetry-allowed.}
\textcolor{red}{The two elements also differ in robustness.  $\Theta$
requires a lattice translation $\bm\tau$ with $\mathbf Q \cdot \bm\tau=\pi$,
which exists only for even commensurate periods.  The helicity mirror $g$
does not require a half-period lattice translation and can remain compatible
with an incommensurate texture.  An incommensurate counter-spiral that retains
$g$ therefore realizes the $\Theta$-broken, $g$-intact sector.}

\begin{figure}[H]
  \centering
  \includegraphics[width=0.78\textwidth]{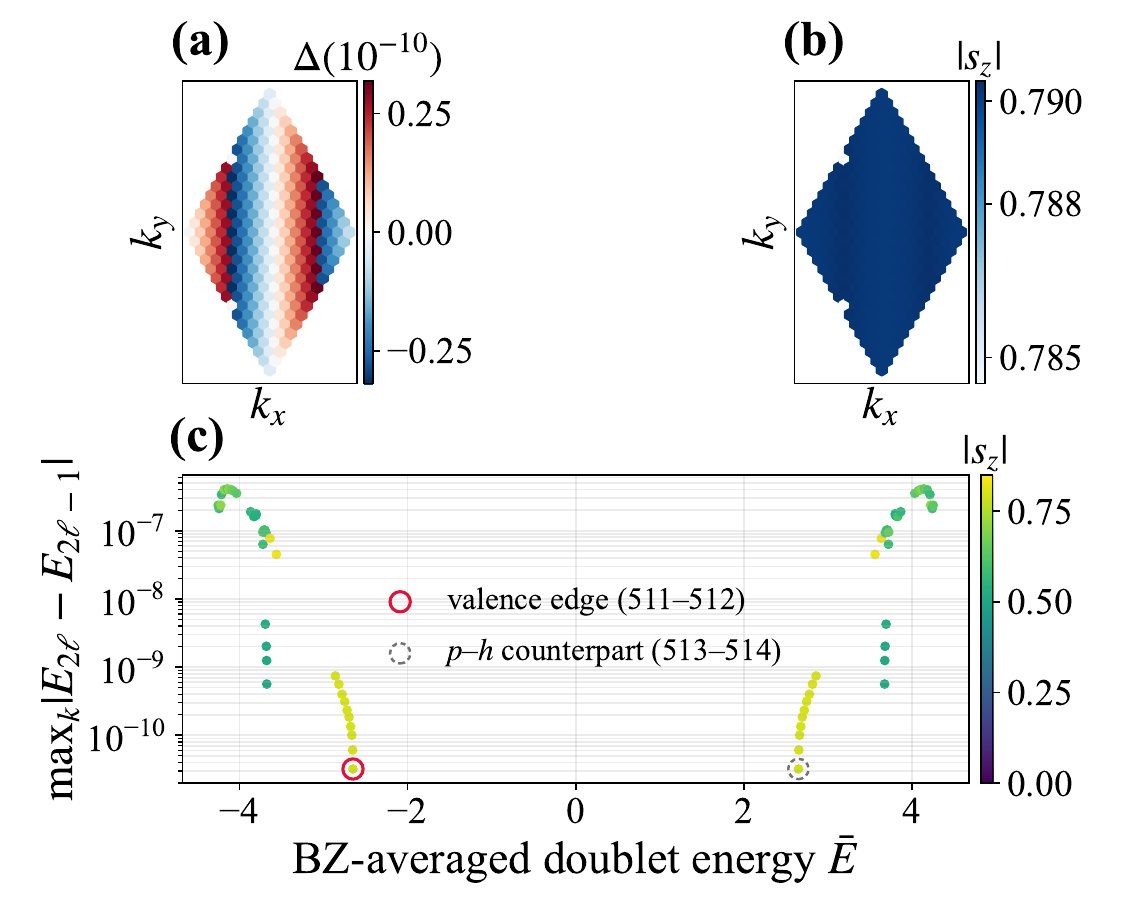}
\caption{\textcolor{red}{Band diagnostics,} $16\!\times\!16$ $\mathbf k$ mesh.
  (a) Signed splitting $\Delta(\mathbf{k})$ of the valence-edge doublet
  $(511,512)$\textcolor{red}{: maximum} $3.2\times10^{-11}\,t$,
  \textcolor{red}{odd under} $\mathbf{k}\!\to\!-\mathbf{k}$ and $M_x$.
  (b) Largest $|s_z|(\mathbf{k})$ of $\Sigma^z$ \textcolor{red}{projected on}
  the same doublet.  The two projected-spin eigenvalues have opposite signs
  and equal magnitudes to the displayed precision.  (c) Maximum internal
  \textcolor{red}{separation} versus \textcolor{red}{BZ-averaged} doublet energy
  for locally resolved two-band subspaces, colored by pair-averaged $|s_z|$;
  only pairs \textcolor{red}{below} $10\%$ of the nearest external direct gap
  are shown (\textcolor{red}{76 of 112 scanned pairs spanning bands 401--624}).  \textcolor{red}{Red circle:} valence-edge
  doublet, \textcolor{red}{suppressed} $\sim\!3\times10^3$ below the
  scanned-window median.  \textcolor{red}{Gray dashed:} particle--hole
  counterpart $(513,514)$, \textcolor{red}{matching to four digits.}}
  \label{fig:headline}
\end{figure}

\paragraph{Quenched splitting, frozen polarization.---}%
At half filling,
the occupied spectrum terminates in a sequence of quasi-degenerate doublets,
with $(511,512)$ at the valence-band edge.
\textcolor{red}{Because the cell spin operator
$\Sigma^z=\sum_{i\in{\rm cell}}\sigma_i^z$ does not commute with
$\mathcal H$ in a noncollinear texture, Bloch states have no definite $z$
spin and a direct
$E_\uparrow-E_\downarrow$ splitting is not defined.  We instead diagonalize
$\Sigma^z$ within each isolated doublet and
define $\Delta(\mathbf{k})=E_{+P^z}(\mathbf{k})-E_{-P^z}(\mathbf{k})$ from the
energy expectations of the opposite projected-spin
partners~\cite{Prodan2009} [Sec.~S3.2].  This reduces to the
usual collinear splitting when
$[\Sigma^z,\mathcal H]=0$~\cite{Hayami2019SpinSplitting,Yuan2020GiantSpinSplitting,Smejkal2022}.}
With this convention, the fully compensated texture,
$\sum_i\mathbf{S}_i=0$, is constrained by $\Theta$ to satisfy
$\Delta(-\mathbf{k})=-\Delta(\mathbf{k})$, which eliminates the
even-in-$\mathbf{k}$ component
\textcolor{red}{$\Delta_{\rm even}=[\Delta(\mathbf{k})+\Delta(-\mathbf{k})]/2$}
exactly.  The helicity mirror $g$ further
requires $\Delta(M_x\mathbf{k})=-\Delta(\mathbf{k})$, fixing the sign
alternation across the mirror line.  The calculated $\Delta$ obeys both
relations
[Fig.~\ref{fig:headline}(a), shown for the valence-edge doublet $(511,512)$]
and is therefore the symmetry-allowed odd channel itself. For the valence-edge doublet,
$\max_{\mathbf{k}}|\Delta|=3.2\times10^{-11}\,t$, more than eight orders of
magnitude below the \textcolor{red}{BZ-averaged} $4.7\times10^{-3}t$ spacing to the neighboring
doublet center.  \textcolor{red}{A scan of 112 adjacent pairs spanning bands
401--624 retains 76 locally resolved pairs; within this set, the valence-edge
separation is suppressed by $\sim\!3\times10^3$ relative to the median}
[Fig.~\ref{fig:headline}(c)].  The conjunction of this quenched spectral
separation with well-defined opposite-$P^z$ partners is therefore a near-edge
property of the counter-spiral.

\textcolor{red}{The near-degenerate valence-edge spectrum resembles that of a
conventional compensated antiferromagnet only energetically.  An orbitally
trivial spin-flip translation cancels the spin-weighted Kubo contributions in
the latter, whereas the helicity mirror here selects a surviving spin-Hall
channel upon doping [Secs.~S3 and
S4.3, Fig.~S5].}

Quenched energy splitting does not make the
near-edge states
spinless.  The
projected spin magnitude remains finite: projecting $\Sigma^z$ onto the
valence-edge doublet gives
$|s_z|\simeq0.7902$, with a $0.75\%$ relative range across the zone
[Fig.~\ref{fig:headline}(b)].
The two partners carry this polarization with opposite
signs, so the doublet remains spin-compensated to numerical accuracy at
every $\mathbf{k}$.
The half-filled occupied projector likewise has $C=0$, consistent with the
symmetry-enforced absence of charge Hall response
[Sec.~S4.1].

\begin{figure}[H]
  \includegraphics[width=0.82\textwidth]{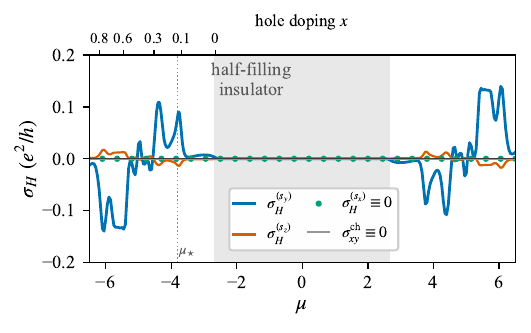}
\caption{Doped spin splitter without \textcolor{red}{spin-split bands}.  Fixed-texture
  $\mu$ sweep for the reference $\mathbf Q_1$ ($M_x$) orientation.
  \textcolor{red}{Dotted line: the representative point
  $\mu_\star=-3.805\,t$ ($x\simeq0.132$) within the metallic window
  $-3.812\,t\lesssim\mu\lesssim-3.799\,t$, where
  $\sigma^{(s_y)}_{H}=0.082\,e^2/h$.}  The helicity mirror
  enforces $\sigma^{(s_x)}_{H}=0$,
  $|\sigma^{(s_z)}_{H}|/|\sigma^{(s_y)}_{H}|\approx\textcolor{red}{0.155}$,
  and $\Theta$ keeps
  $\sigma^{\rm ch}_{xy}=0$.  Gray shading: the half-filled gap; top axis: the
  finite-smearing conversion $x(\mu)$.  $16\!\times\!16$ mesh,
  $\eta\!=\!0.1$, \textcolor{red}{$k_BT\!=\!0.03$,} $J_H\!=\!5$.}
  \label{fig:splitter}
\end{figure}

\paragraph{Spin-splitter transport.---}%
At half filling every linear Hall channel vanishes
[Sec.~S4.2].  Varying the chemical
potential within the gap does not change the electron or hole filling, and hole doping begins only
when $\mu$ enters the valence bands.  In the finite-smearing transport
calculation we convert $\mu$ to hole doping through
$x(\mu)=1-2\nu(\mu)$, where $\nu(\mu)$ is the filling obtained from the same
Fermi occupation used in the Kubo sum of
Sec.~S4.2~\cite{ShiProperSpinCurrent2006,SinovaRMP2015}.
\textcolor{red}{The top axis of
Fig.~\ref{fig:splitter} displays this conversion.}
\textcolor{red}{Both elements continue} to forbid the charge Hall response at
finite doping, while the spin channel becomes active.  Figure~\ref{fig:splitter} shows this evolution
for the reference $\mathbf{Q}_1$ orientation whose $M_x$ helicity mirror is
shown in Fig.~\ref{fig:texture}.

\begin{figure}[H]
  \includegraphics[width=0.80\textwidth]{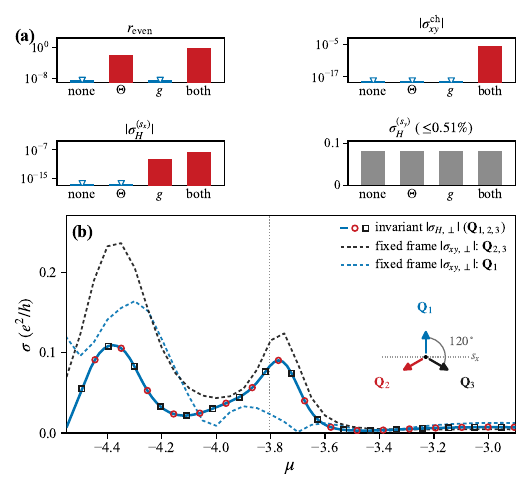}
\caption{\textcolor{red}{Two-element protection and $\mathbf Q$
  reconfiguration.  (a)~Four-sector test at $\mu_\star$ versus the removed
  element(s): $1Q$ potential $V_0=0.03\,t$ removes $\Theta$, imbalance
  $m_{AB}=0.05\,t$ removes $g$.  Blue: channel pinned; red: released.  The
  even channel $r_{\rm even}=\max|\Delta_{\rm even}|/\delta_{\rm
  raw}^{\max}$ opens only when $\Theta$ is removed, the dark-axis channel
  $|\sigma^{(s_x)}_H|$ only when $g$ is removed, and the charge Hall only when
  both are removed.  These three quantities use
  logarithmic scales; symmetry-forbidden blue entries are placed at the
  plotting floor.  The allowed $\sigma^{(s_y)}_H$ (lower right, linear
  scale) stays at $0.082\,e^2/h$
  [Sec.~S6.1].}  \textcolor{red}{(b)}~Fixed-texture $\mu$ sweep for
  the three $\mathbf Q$ orientations.  The invariant $|\sigma_{H,\perp}|$
  collapses onto one curve ($\mathbf Q_1$ line;
  $\mathbf Q_{2},\mathbf Q_3$ markers), whereas the fixed-frame
  $|\sigma_{xy,\perp}|$ (dashed) includes the symmetric part and differs
  between orientations.  Rosette: spin-Hall vectors at $\mu_\star$,
  $120^\circ$ apart, keying the colors; \textcolor{red}{dotted ray: detector
  axis $s_x$, the dark axis of $\mathbf Q_1$.}  Dotted vertical line:
  $\mu_\star$.  $16\!\times\!16$ mesh, $\eta\!=\!0.1$\textcolor{red}{,
  $k_BT\!=\!0.03$}.}
  \label{fig:reconfig}
\end{figure}

\textcolor{red}{Unlike the charge conductivity, the off-diagonal spin
conductivity need not be antisymmetric because its current and drive operators
are different.  We therefore report the rotationally invariant Hall part
$\sigma^{(s_a)}_{H}\!\equiv\!\tfrac12(\sigma^{(s_a)}_{xy}
-\sigma^{(s_a)}_{yx})$ [Sec.~S4.2].
Figure~\ref{fig:reconfig} separately compares $|\sigma_{H,\perp}|$ with the
fixed-frame magnitude $|\sigma_{xy,\perp}|$, which also contains the symmetric
part.}

\textcolor{red}{Bands 445--448---the Fermi doublets $(445,446)$ and
$(447,448)$---cross the Fermi level over the metallic window
$-3.812\,t\lesssim\mu\lesssim-3.799\,t$, with maximal pair separation below
$1.8\times10^{-7}\,t$.  Around the representative point
$\mu_\star=-3.805\,t$ ($x\simeq0.132$), the spin--orbit-free Kubo response is
$\sigma^{(s_y)}_{H}=0.082\,e^2/h$ at $\eta=0.1\,t$.  It is
$N_k$-converged and remains finite over
$\eta=0.05$--$0.15\,t$ [Sec.~S4.2].  The helicity mirror
sets $\sigma^{(s_x)}_{H}=0$, while
$|\sigma^{(s_z)}_{H}|/|\sigma^{(s_y)}_{H}|\approx0.155$.}
This doped regime realizes the \emph{spin splitter}: a transverse spin
current without a transverse charge current, generated here by the Berry
curvature of quasi-degenerate bands rather than by band spin splitting.
\textcolor{red}{Band pairs closer than $0.01\,t$ contribute zero net weight
to this response, which is instead carried by finite-separation interband
matrix elements [Sec.~S4.2].}
Because the allowed spin polarization is fixed by the helicity mirror that
defines the SSG operation, this response is the transport manifestation of the
texture altermagnet.

\textcolor{red}{Carrier back-action is bounded variationally: the
counter-spiral stays below the optimal ferromagnet and carrier-induced
N\'eel state over $0\le x\le0.30$ for $|\Gamma'|S^2/t>1.51$
[Sec.~S4.4].}

\paragraph{Two-element protection.---}%
\textcolor{red}{The $\Theta$--$g$ allocation is established by separately
breaking the two elements on the same bands and mesh
[Fig.~\ref{fig:reconfig}(a); Sec.~S6].  The
spin-independent $1Q$ scalar potential, namely $V_i=V_0\cos(\mathbf{Q}\!\cdot\!\mathbf{r}_i+\varphi_{s(i)})$
with sublattice phases $\varphi_A=0$ and $\varphi_B=\pi$ chosen to keep it
even under $g$, breaks $\Theta$ alone, since
$\mathbf{Q}\!\cdot\!\bm{\tau}_{1/2}=\pi$ flips its sign under the half
translation.  A uniform sublattice imbalance, $+m_{AB}$ on $A$ and
$-m_{AB}$ on $B$, preserves the half translation but is odd under the
sublattice-exchanging mirror, removing $g$ alone.  Each perturbation
separates the Fermi doublets.  What distinguishes them is whether
$\Delta_{\rm even}$ opens with it.  The $1Q$ potential at $V_0=0.03\,t$ raises
$\max|\Delta_{\rm even}|$ to $1.1\times10^{-2}$ of the raw band separation
$\delta_{\rm raw}$, whereas the imbalance $m_{AB}=0.05\,t$ keeps the same
ratio at zero while separating the bands twice as far:
the exact even-channel constraint belongs to $\Theta$, not to $g$.  The
charge Hall remains zero in either single-break sector and
reaches $2.81\times10^{-6}\,e^2/h$ only when both elements are removed.  The
$s_x$ zero follows $g$ alone, and the allowed $s_y$ response varies by at
most $0.51\%$ across all four sectors.}

\textcolor{red}{To test the relativistic continuation of $g$, we add a
mirror-compatible intrinsic honeycomb spin--orbit term of strength $\lambda$
to the $\Theta$-breaking $1Q$-potential configuration above, which retains $g$.
At $\lambda=0.1\,t$, the charge-Hall and $s_x$ spin-Hall channels remain zero,
while the magnitude of the allowed $yz$-plane spin-Hall response decreases by
$6.6\%$ [Sec.~S6.2].}

\paragraph{Reconfigurability.---}%
Unlike a crystal-fixed altermagnet, the non-translation operation here follows
the ordered wavevector.  The three degenerate $\mathbf Q$ orientations are
related by the combined spin--lattice $C_3$ symmetry of the $\Gamma'$ exchange
and carry correspondingly rotated helicity mirrors
[Sec.~S1.3].  Selecting $\mathbf Q$ therefore rotates the
allowed in-plane spin-Hall polarization by exactly $120^\circ$, while leaving
its magnitude invariant.  At $\mu_\star$, the three orientations share
$|\sigma_{H,\perp}|=\textcolor{red}{0.082\,e^2/h}$.  Figure~\ref{fig:reconfig}\textcolor{red}{(b)} shows this
invariance directly: across this regime the three orientations fall on a
single $|\sigma_{H,\perp}|(\mu)$ curve, while the full
fixed-laboratory-frame magnitudes $|\sigma_{xy,\perp}|$ differ between
orientations, and the rosette displays the three spin-Hall vectors separated
by $120^\circ$.  Experimentally, for a fixed spin-detection convention, a
$y$-directed drive with spin-current flow detected along $x$ gives
$\sigma^{(s_a)}_{xy}$, while exchanging the drive and flow axes gives
$\sigma^{(s_a)}_{yx}$. Hence their half-difference isolates
$\sigma^{(s_a)}_H$ [Sec.~S5.1].

\paragraph{Discussion.---}%
\textcolor{red}{The counter-spiral realizes charge-Hall-free spin-splitter
transport in the quenched-splitting regime.  Its symmetry constraints
are divided between two elements: the antitranslation $\Theta$ forbids
even-parity band splitting, while the helicity mirror $g$ fixes the
spin-polarization and dark axes.  Either element independently enforces the
charge-Hall zero.  Selecting one
of the three degenerate orientations rotates the allowed and dark spin axes in
exact $120^\circ$ steps while preserving the charge-Hall zero.}

\textcolor{red}{The self-organized realization studied here requires
bond-anisotropic exchange that stabilizes the counter-spiral and carriers whose
back-action preserves it [Secs.~S1.4 and
S4.4].  Weak directional anisotropy, strain, or boundary
pinning provide possible preparation-stage controls for selecting $\mathbf Q$.
Counterrotating order in the Li$_2$IrO$_3$ polytypes provides a material
precedent for the texture and motivates the search for commensurate itinerant
analogs~\cite{Williams2016alphaLi2IrO3,Biffin2014betaLi2IrO3,
Biffin2014gammaLi2IrO3,KimchiColdeaVishwanath2015}.  High-spin van der Waals
honeycomb magnets are complementary starting points for the $\Gamma'$--$D$
route~\cite{KitaevReview2024}.}

\textcolor{red}{Photonic platforms already engineer altermagnetic symmetry in
static dielectric patterns fixed at
fabrication~\cite{Kim2025PhotonicAM,Qiu2026OrbitalPhotonicAM,Cao2026MirrorPhotonicAM}.
Programmable photonic meshes and topolectrical circuits instead encode the
site-resolved single-particle
Hamiltonian~\cite{On2024ProgrammablePhotonics,Lee2018,Wu2022} and can be
rewritten in situ.  A $64$-mode lattice suffices: the minimal $L\!=\!4$
supercell ($32$ sites) retains both SSG elements and reproduces the dark-axis
zero, the charge-Hall zero, and the exact $\pm120^\circ$ reorientation
[Sec.~S5.2].}

\paragraph{Data availability.---}%
\textcolor{red}{Numerical data and analysis scripts supporting this work are
available from the corresponding author upon reasonable request.}

\begin{acknowledgments}
This work was supported by the National Key R\&D Program of China (Grant No. 2022YFA1402704), by the National Natural Science Foundation of China (Grants Nos. 12274187, 12304176, 12247101), by the Shenzhen Fundamental Research Program (Grant Nos. JCYJ20220818100405013 and JCYJ20230807093204010) and by the Natural Science Foundation of Jiangsu Province (Grant No. BK20241929).
\end{acknowledgments}

\bibliography{refs}

\end{document}